# WORK SHARING AS A METRIC AND PRODUCTIVITY INDICATOR FOR ADMINISTRATIVE WORKFLOWS

Charles Roberto Telles[1]

[1]General Governance. Secretary of State for Education and Sport. Paraná. Brazil.
charlestelles@seed.pr.gov.br

**ABSTRACT**

Defining administrative workflow events as a nonlinear dynamics that assume a random ordered or disordered growth rate of information processing, a method has been proposed for large-scale administrative systems that structures hybrid system variables (continuous or discrete) as iterated and attracted to a fixed-point event at which for all possible metric spaces solutions, the modeling of variables from Lyapunov exponential stability point of view allows the projection of system performance to be oriented, that is, the relationship between the number of agents and the number of administrative services within an administrative workflow environment.

## INTRODUCTION

A mathematical method was developed to generate an indicator on the productivity of administrative workflows. The proposed methodology is only suitable for large-scale administrative activities that have a wide range of activities as well as the number of agents that perform them.

The indicator can roughly and steadily reveal the definition of productivity as a state of endosectoral (among internal sector agents) and exosectoral (among external sector agents) administrative services sharing. The indicator can be used by public and private managers to measure human resource efficiency in proportion to work requests of all administrative services occurring in a given sample. This definition differs from traditional or differentiated key performance indicators (KPI), such as working hours, medical certificate, per capita productivity, among others [1,2,3,4].

The main approach for creating an indicator for productivity analysis, is to be able to extract labor demands and workers amount proportionality within workflow environments where discretization and computerization is not fully supported and/or available. This condition causes the work activity to present uncountable flows of information processing leading the workflow system to an uncertain predictive production. This empirical version of analysis in this article investigates the Regional Education Centers of the State of Paraná (NRE) of the Secretary of State for Education and Sport of Paraná (SEED), Brazil, where large amount of administrative services as well agents plays the role of having as the object of production, the public education administration. In this way, the proposed metrics defines how efficient the operationalization of NRE work is in terms of discretization and computerization, and allows to verify over time whether state actions and / or policies have improved the operationalization of administrative work or are as rules that plaster / overload system.

The elaborated methodology focus in an analysis of the administrative workflows and the flow of processes occurring on these Regional Education Centers of the State of Paraná in relation to the number of existing agents. As an extension to the administrative work of the SEED, the NRE correspond to the linearity of management from education policies in the state. In this sense, by providing better management of NRE, the performance of these educational policies also becomes more efficient and effective.

The main purpose of this research is to enable innovative policy management to unburden NRE work and enable human resource allocation for more emerging tasks, as well as boost NRE productivity in their pedagogical, administrative, control, monitoring, field activities and policies of the Paraná State Government.

Methodologies such as descriptive statistical analysis with descriptive graphs, tabular description and linear regression were used to compose the analyzes ($R^2$) [5] associated with exponential function.

The collected data, through questionnaires, were obtained observationally and qualitatively and later transformed into quantitative samples by compiling the samples. It is important to realize that interviews, the method in which data were collected, may have inaccuracy caused by factors related to the subjectivity with which each agent performs an activity, and thus generates labor productivity. Similarly, cognitive issues and other nebulous factors (systems randomness level) may lead to inaccuracy of the collected data [6,7] however, the quantification of the data allows us to verify the frequency with which a particular event investigated should or should not be considered as objective, as well as the subsequent general quantification (result of collections from all NREs) allows to refine the most objective data within a composite population sample of a variance that permeates the objective and subjective universe of cognitive realities.

Also, some data were not collected because it was not available at the research site or because there was no satisfactory answer regarding the search parameters. However, this did not invalidate data collection since the number of samples and collections were numerically and proportionally significant for the purposes for which this research was performed.

After refining the samples, the information collected was compared to each other generating descriptive statistical results for the interpretation of the events (descriptive graphs) and at this point, it was possible to elaborate an indicator based on exponential functions as an approximation of the agents' potentiation effect (sharing works) in an attempt to cover the large existing requests of administrative services and the lack of computerization and discretization of administrative massive routines.

**METHODOLOGY**
1. **Data collection**

This methodology was created based on data collected in 2019 in the Dois Vizinhos, Francisco Beltrão and Pato Branco Regional Education Centers (administrative units), and the parameters adopted for data collection are related to the number of administrative services, frequency with which they are performed (considering their beginning and ending), number of on-site agents, and efficient productivity interactions (work sharing), as exemplified by Table 1 (the full table is available from the Author Affiliate General Board).

Table 1. Questionnaire used to collect data on administrative process flows in NRE.

| Nº | SECTOR: | AGENTS QUANTITY | WORK FREQ. |
|---|---|---|---|
| 1a | **OMBUDSMAN** | | |
| | Ombudsman's Inquiry | | |
| | SIGO (work related system) | | |
| 2b | **CHIEF** | | |
| | Secretary | | |
| | Communication Advisory | | |
| | Chief Advisor | | |
| | Agent working hours | | |
| 3c | **HUMAN RESOURCES** | | |
| | PR social security | | |
| | Paraná Education Administrative Unit | | |
| | Citizen attendance service | | |
| | Agent career promotion / progression | | |
| | GR Billing (improper benefit) | | |
| | E-protocol and archiving | | |
| | Citizen requirements / resources | | |
| | Work accident protocol | | |

A total of 17 large sectors of work were analyzed, containing in total administrative services in each sector, 158 distinct attributions related to the work of the NRE, Secretary of State for Education and educational establishments.

In table 1, the work frequencies are divided into high, medium and low, and the qualification criteria among the categories is defined by the agent's response regarding how long each day of the week in each work request, the agent is busy with the activity, which can occupy almost every working day (high), or rarely (low) or a few days a day (average).

Also in table 1, it is divided with numbers the main sectors of the NRE (17), which historically have been reified by the public institution and state policies. Along with the main sectors, follows the demands of each sector (total of 158). Associated with the sectors and their demands, to measure the efficient interactions of productivity, it was verified whether the agents of the sector itself share or take turns on a certain demand for work, or even if other agents from external sectors also play the same role in terms of an efficient interaction of worker's productivity. This possibility of sharing work aiming at production efficiency was analyzed and described with the help of the letters indicated by a, b and c, which are on the right side of table 1 for each of the main sectors represented by numbers 1, 2 and 3.

The importance of this parameters of analysis was to identify whether the discretization and computerization effects of reducing workload for each agent can be tracked over the years as far as state policy can improve administrative workflows, leading agents to a better performance of executing an activity in terms of lowering time, costs, precision reaching and resources and turning possible to achieve higher standard of work allocation and productivity within the equation "agents quantity X requests quantity X number of administrative services".

From the collected data it was possible to identify important characteristics about the productivity of the agents, highlighting mainly the number of NRE schools in relation to the number of agents in each NRE sector as well as the total number of agents per NRE and administrative services by sector. The NRE administrative services by sector (workflows) can be seen in figure 1.

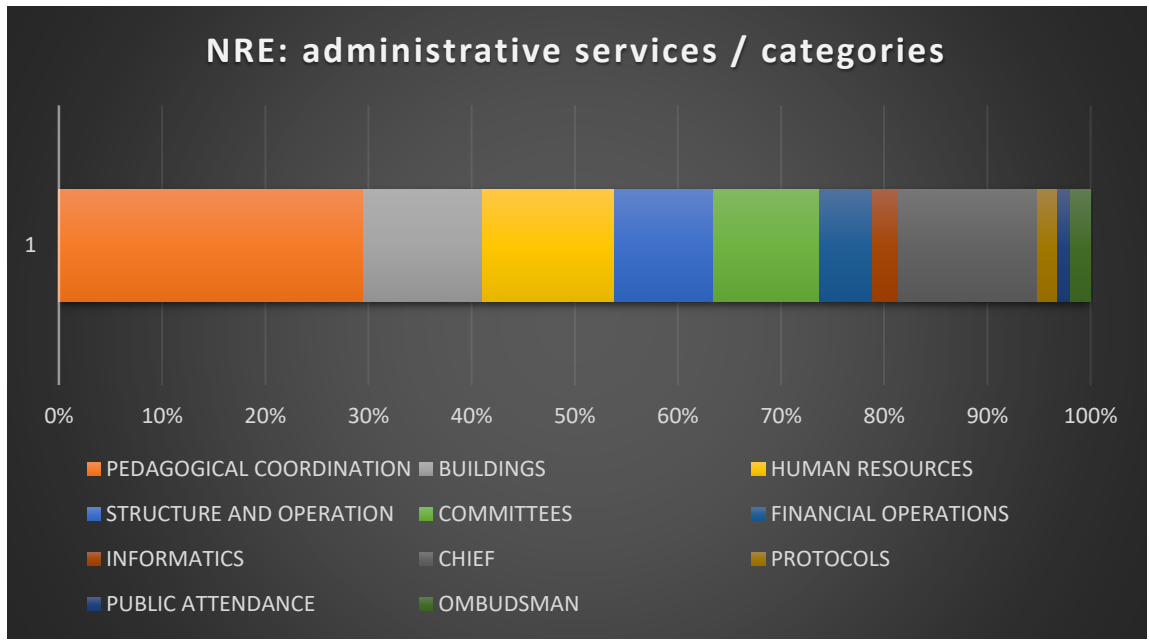

Figure 1. In general, excluding smaller administrative services or those that were not erroneously recorded in this research, the sectors of an NRE present the following official distribution of work in proportions for each sector.

From figure 1, the main problem found was the questions of how to manage human resources and administrative services that are in dimensions in which the nature of the work performed cannot always be known objectively, generating the notion that some variables of workflows are either discrete in nature (observable and easily quantifiable) or continuous in nature (difficult to recognize and manipulate). In this sense, one main question that is the purpose of this research arises with the statements: how much agents can affect productivity (are necessary) in a given sector when both types of work nature (discrete and continuous variables) are occurring? And is it possible to reach an efficient management point? In order to address this questions, a quantitative approach to measure productivity within a nonlinear dynamical event (information processing as a complex adaptive system (CAS) [8,9]) is required.

Two main points that affect productivity in a workplace that consists in discretizing activities and enhancing production through shared ways of producing from existing human resources were used to estimate productivity towards human resources and demands of administrative work in the given circumstance described at introduction section.

## 2. Methods of Analysis
### 2.1 Extracting a Productivity Indicator for NRE

**Definition.** Taking as variables that promote productivity, the input consisting of the number of agents in the place, the endosectoral collaboration (agents that help each other within the sector itself) and exosectoral collaboration (agents that help each other across distinct sectors), total labor requests (number of schools to be attended) represent all these variables the actual numbers of $y(a)$ where in effect of an exponential (increasing or decreasing) function of the type $f(x) = ab^x$, the function can indicate how the input of the equation becomes proportional to the output as an index of system productivity, and it is possible for the manager to manage the input as the results are collected and observed over the years [10] (represented by $b$) in order to achieve a less asymptotic curvature of function [11].

The later use of the coefficient of determination $R^2$ from the formula $r = \frac{\sum(x-\bar{x})(y-\bar{y})}{\sqrt{\sum(x-\bar{x})^2 \sum(y-\bar{y})^2}}$, describes from each component of the exponential function the proportionality between the bases and their powers in a continuous process of distribution of the variables of each event to each other, considering this as the multiplication effect of the exponential function equation that is represented as $f(x) = ab^x$ (in graphic form $f(x) = yx^e$), where in the relationship between the variables number of agents, endo and exosectoral effects with the administrative services of educational institutions, each $a$ and $b$ represent respectively the value of $y$ for each discrete variable and the value of $x$ for each procedural relation (time) of the stream of events (continuum) in potentiation. This function can be represented as $f(x) = ab_1^{x_1}.ab_2^{x_2}.ab_3^{x_3},\ldots,ab_n^{x_n}$, in which $x$ is a constant of non-numerical nature in concrete terms of analysis, but that for the production of a function, consider $x$ as a frequency rate $e$ (Euler number) with which there is the growth or reduction of events assuming for this that for all the variables in question, the maximum effect of potentiation is approximate to reality, not being possible to really know each particularity of the real life. This does not affect the observation of the phenomenon since the extraction of an indicator for productivity should take into account approximately the first causes that lead one NRE or another to have more productivity from concrete variables such as the number of agents, efficient mechanisms for work sharing and the total number of schools.

Two points are important in using a productivity indicator for NRE, which is to observe the $R^2$ of each sample and at the same time, as another source of system information, the curvature of the exponential function calculated from a constant defined by a derivative such as $x = \frac{d_{a_i}}{d_t} b^x = f_i(t, a_1, a_2, \ldots, a_n) b^x$, where $i = 1, 2, \ldots, n$, enabling the observation of the power of each iterated event as an effect that causes consecutive actions on the dynamics of the event in sequence. This dynamic reflects very objectively the behavior of events in interaction and continuous iteration when the discretization and computerization effects over the years are empirically true via data collection, which in the exponential function can be observed by the curvature of the generated function and can empirically reflect in the high growth rate or reduction of overlapping effects with an initial condition defined by $a_i(t_0) = a_{i0}$. For the $R^2$, it is possible to observe the ratio between the function components separately, which allows you to view the ratio of an input and output of productivity. As for the curvature of the exponential function, it is observed how the iterated and continuous event changes over time, reflecting whether or not the work requests output has excessively greater weight in the input balance reflecting precarious or primitive human resources allocation and at the same time the workflow technological and/or scientific modernization.

Assuming that administrative workflows cannot always be observed regarding agents best proportion towards work requests, the nonlinearity of events in terms of production sharing or even production per capita are events which, although described exponentially as an approximate indicator of the process, it does not reflect the mode of production of the site, and other mathematical tools are needed for this later description. Following this path, this proportionality between agents and requests of the phenomenon may or may not be possible due to nonlinear nature of events when cognitive processes are also the main component of production [6,12]. This can be considered as a first limitation of the method proposed that is the inherent complexity of a workflow when it is composed of human-machine n-dimensions.

Besides it occurs as a limit, the emergent phenomena still can be observed by the indicator proposed in terms of human resources variance through years or machine environment interfaces.

This finding leads us to one of the main points to be discussed in this research, already mentioned, which deals with the discretization of events and enhancement of labor productivity, which parameters are directly caused and affected by two functions of linear and nonlinear characteristics simultaneously, which are the processing of information by the agent and the operationalization of activities. These two points will not be deeply addressed, but are contemplated in the following analyzes by defining each of them from an observational and stationary perspective of the phenomenon (statistics) not to say about the mechanism of operation that causes the phenomenon to exhibit one or other outcome (probabilistic or deterministic nature of administrative services and processing by agents).

### 2.2 Modeling of administrative asymptotic instability events as an exponentially iterated function under Lyapunov exponential stability

Another essential point in formulating an indicator for NRE productivity is that there is an asymptotic imbalance in the solutions for the entire administrative event itself, considering all the work, routine and agent CAS dynamics. If considering the system by its natural expression, it is possible to know mathematically that there will never be a viable unique solution to the question of the ratio between the number of agents and NRE administrative services. In general terms, asymptotic instability only reaches stability when the number of administrative services is always greater than the number of agents (natural form of the event in mostly administrative public services). From this, the proposition that an exponential function may represent a reduction in the metric of possible spaces for solutions around the given fixed point (Lyapunov stability) [11], which is the number of educational establishments, can serve positively and/or forcibly to prove that an indicator based on these assumptions will have greater management validity than other approaches that do not consider the system as non-stationary and possessing high complex initial perturbations (multiple solutions).

**Theorem 1.** Consider that the variables $a$, $b$ and $t$ are defined within a maximum metric space $\delta$ defined by the interaction and iteration between the number of agents and the number of schools, it must assume solutions $\varphi$ lower or equal to $\varepsilon$. Having $n$ possible initial trajectories $X(t_n)$ as input data, the solutions $\varphi(t_n)$ should be contemplated with a view to better efficiency of the public expenses as an exponential function thus permeating Lyapunov's exponential stability [11] as a goal of public policies to be adopted.

**Proof of the theorem 1.** Let's assume that a trajectory of solutions $X(t_1)$ input to one of the components of the function represented by $f(x) = a_1 b^x$, a number of agents must necessarily satisfy the solution for an output of $\varphi(t_n)$ of a number of maximum establishments of the system represented by the function component as $f(x) = a_n b^x$. In terms of empirical description of this type of event, for every number of agents allocated to a particular NRE, there is a demand for administrative service such that this number should compensate for the number of establishments in the NRE's geographic region.

This obligation leads the phenomenon to necessarily present an asymptotic stability of Lyapunov defined by $X(t_n) - \varphi(t_n) < \delta$, so the system already presents by natural government induction an attempt to control $\delta$ on the large-scale administration of massive public administrative events.

The organization of the system, defined by an asymptotic stability, assumes, by empirical nature of the system (discretization and computerization), a potentiation (Lyapunov exponential stability) in which the phenomenon for most public spaces can be represented as a desired phase state as proposed in this research as a modeling of public administration by work sharing as a metric and productive indicator, like $X(t_n) - \varphi(t_n) < \delta$ where $|X(t_n) - \varphi(t_n)| \leq \alpha |X(t_0) - \varphi(t_0)| e^{-\beta t}$, being $\alpha$ and $\beta$ the empirical unknown parameters of the exponential function.

Q.E.D. ∎

For system management purposes, it is possible to view the possibility of administrating system discretization, computerization and monitoring (NRE) in order to be able to predict and validate the exponential function as a valid indicator of nonlinear systems (NRE) where the results were defined by the amount of work sharing effect. This modeling of systems by an assumption of exponential stability such as Lyapunov allows us to observe the phenomenon with a maximum metric limit of possible solutions to be defined by the conception of the exponential function itself seeking to adapt the empirical events to the search for a metric to be achieved, which in turn represents the productivity that involves human resources and the number of educational establishments.

All actions that fall between the administrative process flows that result in the modification of variables that affect the system for the agent's work must be exponentially a system convergence output, since they imply that variables, for example, such as the time, can reduce the work execution of an agent allowing it to be allocated to another activity, thus allowing the system potentiation effect to be generated, which is something that can be identified by the indicator proposed in this research. This evaluation leads to a possible policy making where indicator will or not prove which empirical experiments performed by a government present the desired growth of exponential productivity. More than an indicator the tool can also serve as inductive procedure within complex environments such as commonly found in public administration.

Another point to be considered for productivity analysis in the relationship between number of people and administrative work periodicity is the time factor. Empirically, many events are difficult to predict, which also affects the possibility of identifying a possible convergence in exponentially proposed solutions. Although some periodicities, due to their high frequency rate, are more easily observed, other administrative services periodic expression will appear to have a false impression of ineffectiveness. This is due to a false effect generated by the time factor, in which an exponential or asymptotic stability, empirically, can only be observed over large time scales. However, if it is not possible to observe, it does not mean that it will not in fact produce convergence to the fixed point (productivity definition), or in other words, be outside the proposal of conceiving large-scale administrative events with the exponential stability method as far as the empirical basis of the method proposed are being considered for all investigation as inductive and tool-oriented knowledge database.

**2.3 Discretization and computerization rate and indicator boundaries**

Following similar discussion at item 2.2 of methodology section, despite of productivity indicator accuses a growth rate, it does not reflect the entire dynamics of workflows that follow nonlinearly, that is, any input of work that is passed on to the NRE, the internal or external organization of a sector to increase frequency or production capacity is not always linear, either because of the cognitive questions of agents that do not deterministically operate the flow of information, or because the nature of services does not have full endo or exosectoral possible interaction. Thus, there is no way to define a numerical rate (other than $e$) that represents a production exponential that can be used, in its numerical form, as an indicator of productivity within a context in which the system is highly random, so it does not generate predictive time series closer to reality.

The production rate $x$ in real exponential mode will not be the object of study to be used for analysis purposes, requiring for the object of study of this nature methodologies that can describe the productivity events in continuous variables from probabilistic definitions as can be define them by the productivity equation $i = Prob\left(\frac{T^i}{I}\right).Prob\left(\frac{T}{I}\right) = m + i$, where $i$, $T$, $I$ and $m$ respectively represent productivity, time, information and probability distribution of $i$, added to another event $i$, if it occurring continuously [14]. However, these approaches to the functioning of administrative workflows will not be addressed in this research in the sense that these observations become one more component of observation of the proposed method.

In contrast, the productivity equation [13] in its form $i = Prob(I^i > n).Prob\left(\frac{T}{I}\right) = \int_P^{I^{i>n}} f_I\left(\frac{T}{I}\right) dI^i$, reflects the particularity of the nonlinear systems of NRE in the operationalization and standardized ability of agents to perform their assignment activities. This particularity acts as a projection of the system to reduce randomness allowing an evolution of information flow and processing to add to the concept of work sharing productivity. Discretization, computerization, and information accuracy events in administrative workflows achieve greater efficiency of productivity and they can be tracked inductively by indicator proposed.

However, one may wonder why defining the event with exponential function if the rate that legitimizes the graph of the function is not viable to conceive the nature of the event? It is important to note that as a stationary analysis of the event, the event is conceived in its exponential function as a dynamic between the values of the real numbers that make up $y$ in $b$, and that in the number of agents and other real variables, the function iterated in $x$ develops potential expression chaining as a multiplication of the real values such that the proportionality between the input and output must necessarily be a closer description of the productivity events as an enhancement between input and output, which is a fact occurring in the work routine according to data collection carried out in 2019. The graphical function allows us to see how potentiation and continuum between variables occurs and later what proportion there is between input and output in the iterated and potentiated function. Thus the growth rate or reduction of a potentiation in function which is also defined as a natural logarithm $b_x \log_e b$ (e= 2.71828...) [14] may be useful only as a stationary observation of the system to evidence events in an iterated and linear manner, but the rate if used to describe the mechanism of operation of events, as already mentioned in the previous paragraphs, the investigated event does not necessarily behave while a constantly iterated, multiplied, potentiated, or even linear event approaching how otherwise it might be elegantly observed.

Importantly, there is a difference between establishing a stationary indicator to observe an event from nature and saying that the event in its formation is exactly the method of observation.

Thus, the exponential function can serve as a stationary indicator of productivity in NRE, but this also does not define the form of productivity that occurs by more probabilistic and nonlinear natures and at the same time linear and discretized events.

The topic referred to in this last paragraph will be presented in the discussion section of this research, in which a distinction will be made between the use of the indicator as an observation of productivity in terms of operational efficiency and other causes that allow greater efficiency and effectiveness in the execution of work allowing greater collaboration between agents and / or allocation of multiple per capita administrative services index, and the use of the indicator as somewhat limited to the sample population margin or data collection inconsistency, which may lead to analyzes flaws depending on the object to be investigated.

**RESULTS**

**3. Descriptive statistics and the investigative boundary caused by nonlinearity of events**

It is important to point out that the objective of this research was to elaborate a method of analysis of administrative units, bringing mathematical concepts closer to the flow of administrative processes, but especially highlighting which parameters are of great importance to enable the public manager to make a decision regarding the randomness of data that commonly contemplates these administrative work scenarios.

From the data collection for three NRE samples from the regions represented by the Regional Education Centers of Dois Vizinhos, Francisco Beltrão and Pato Branco (Figure 2), in the first semester of 2019, it was possible to visualize the number of agents in each work sector of an NRE, as shown in figures 3, 4 and 5.

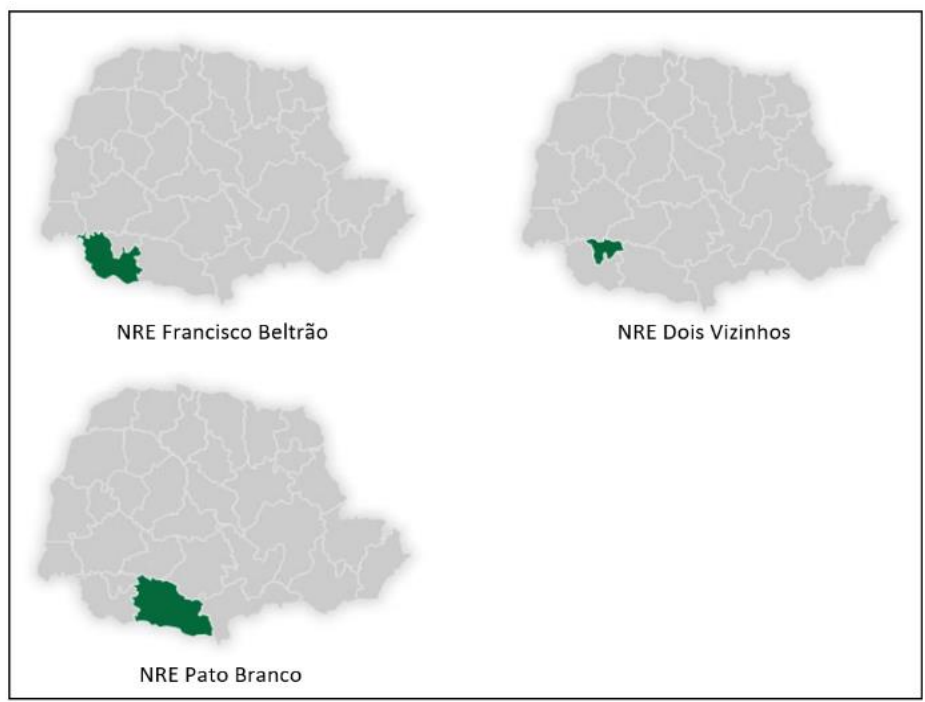

Figure 2. NRE regions investigated at southwest of State of Paraná.

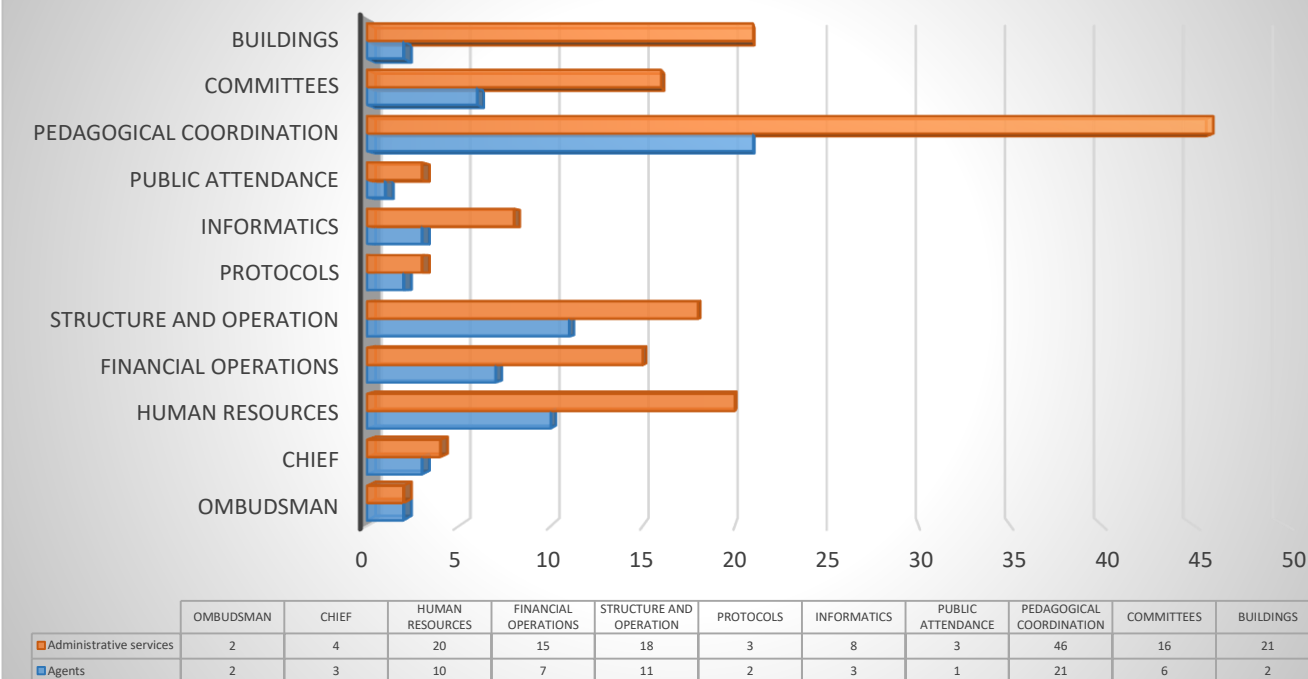

Figure 3. Dois Vizinhos NRE, number of agents and administrative services from Table 1.

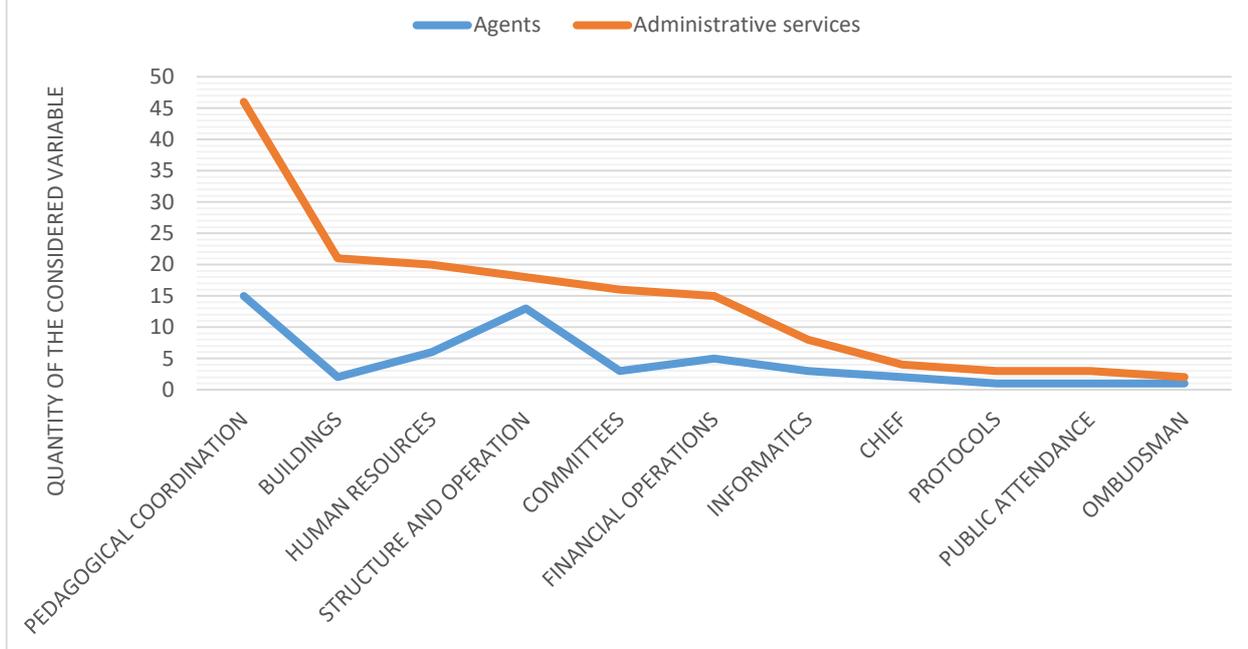

Figure 3a. Dois Vizinhos NRE, number of agents and administrative services from Figure 3.

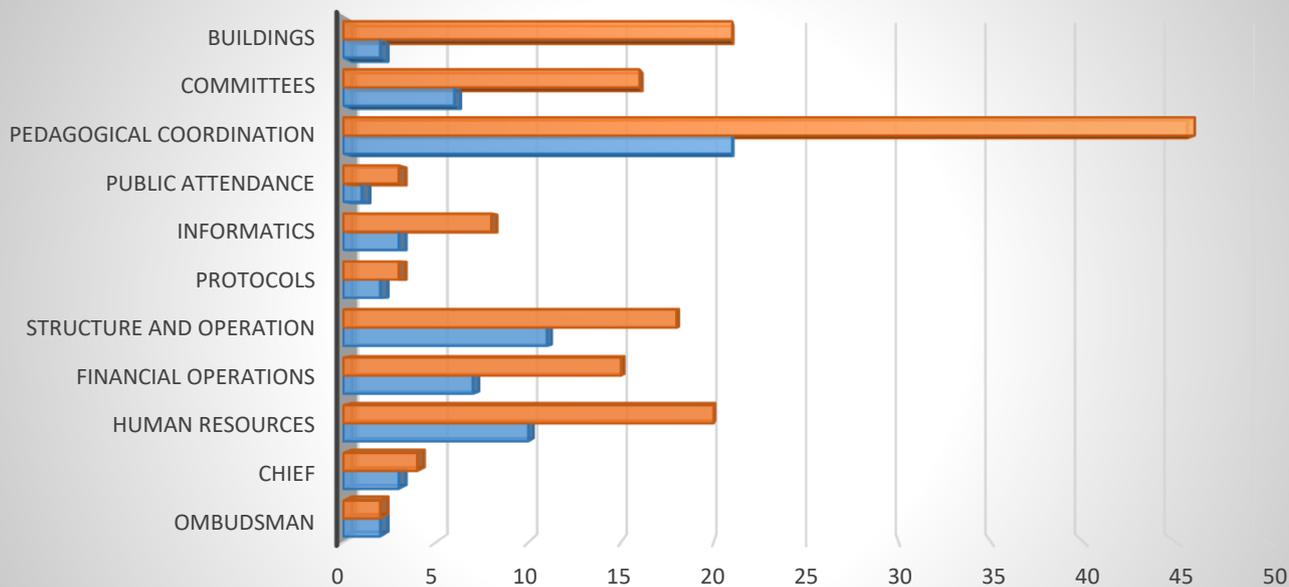

Figure 4. Francisco Beltrão NRE, number of agents and administrative services from table 1.

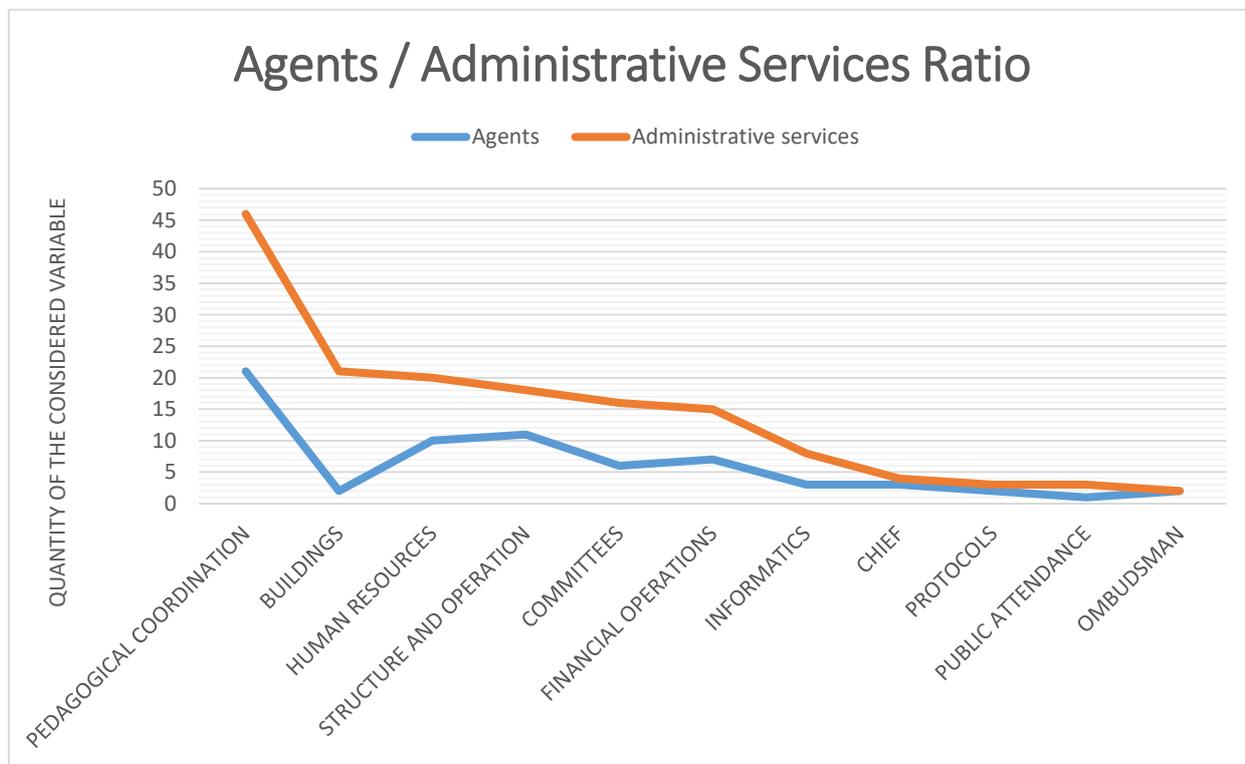

Figure 4a. Francisco Beltrão NRE, number of agents and administrative services from Figure 4.

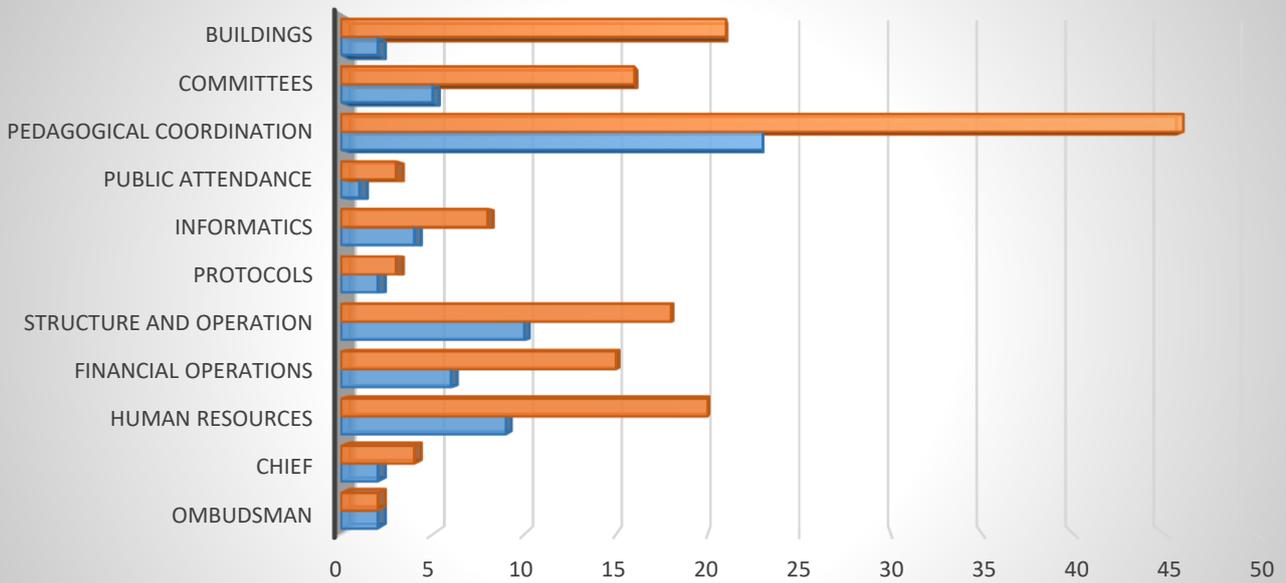

Figure 5. Pato Branco NRE, number of agents and administrative services from Table 1.

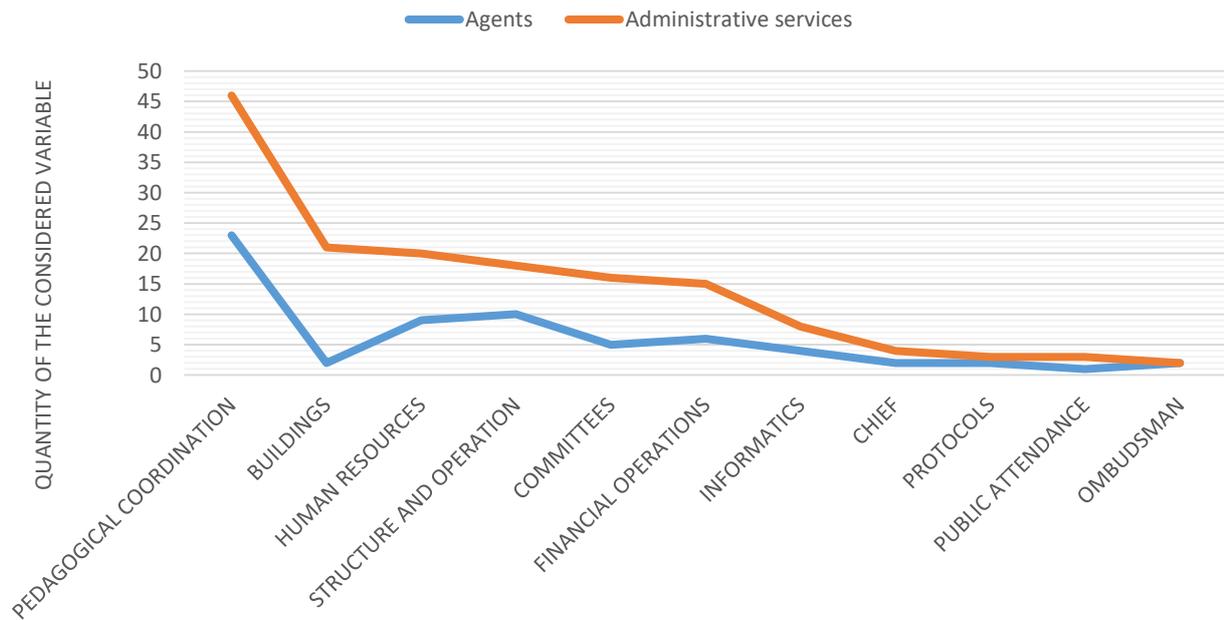

Figure 5a. Pato Branco NRE, number of agents and administrative services from Figure 5.

Otherwise, the descriptive graph of figures 3, 4 and 5 can be viewed in terms of proportionality of the number of agents per sector by figures 6, 7 and 8.

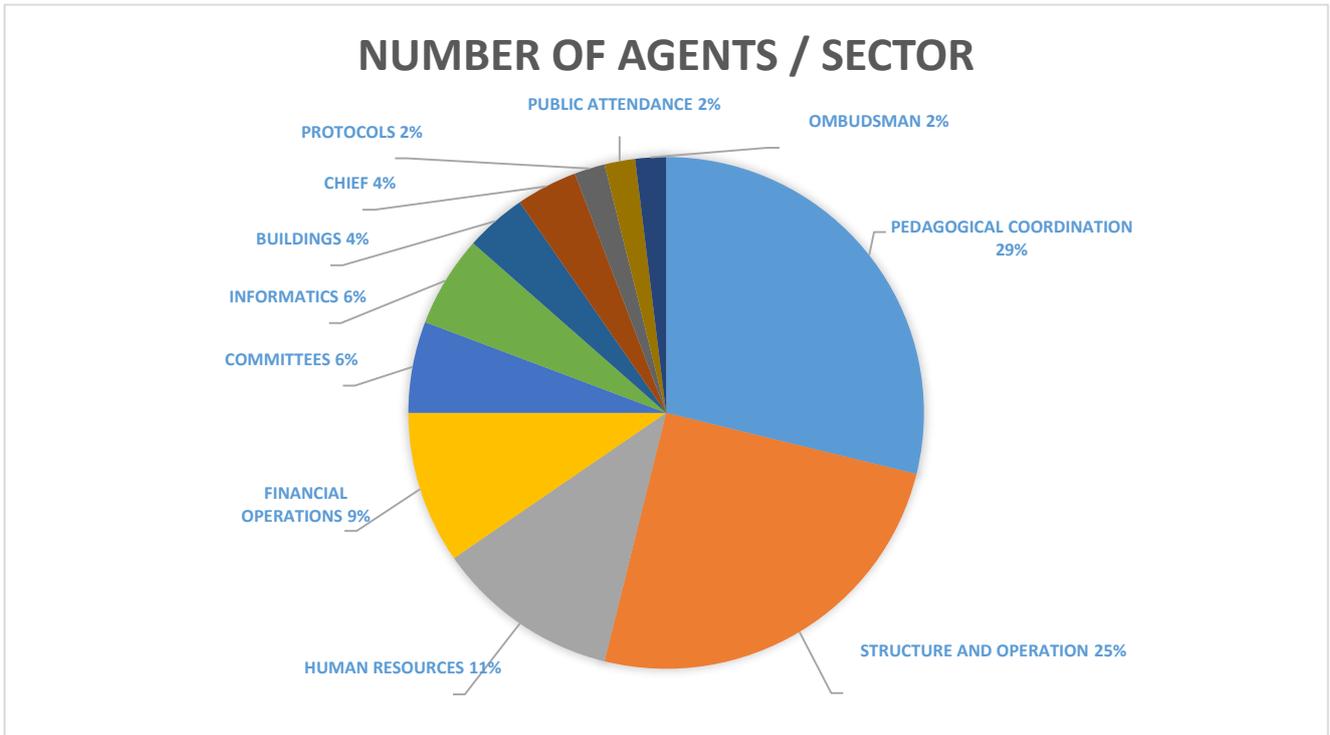

Figure 6. Two Neighbors NRE, proportion of agents by work sector from Figure 3.

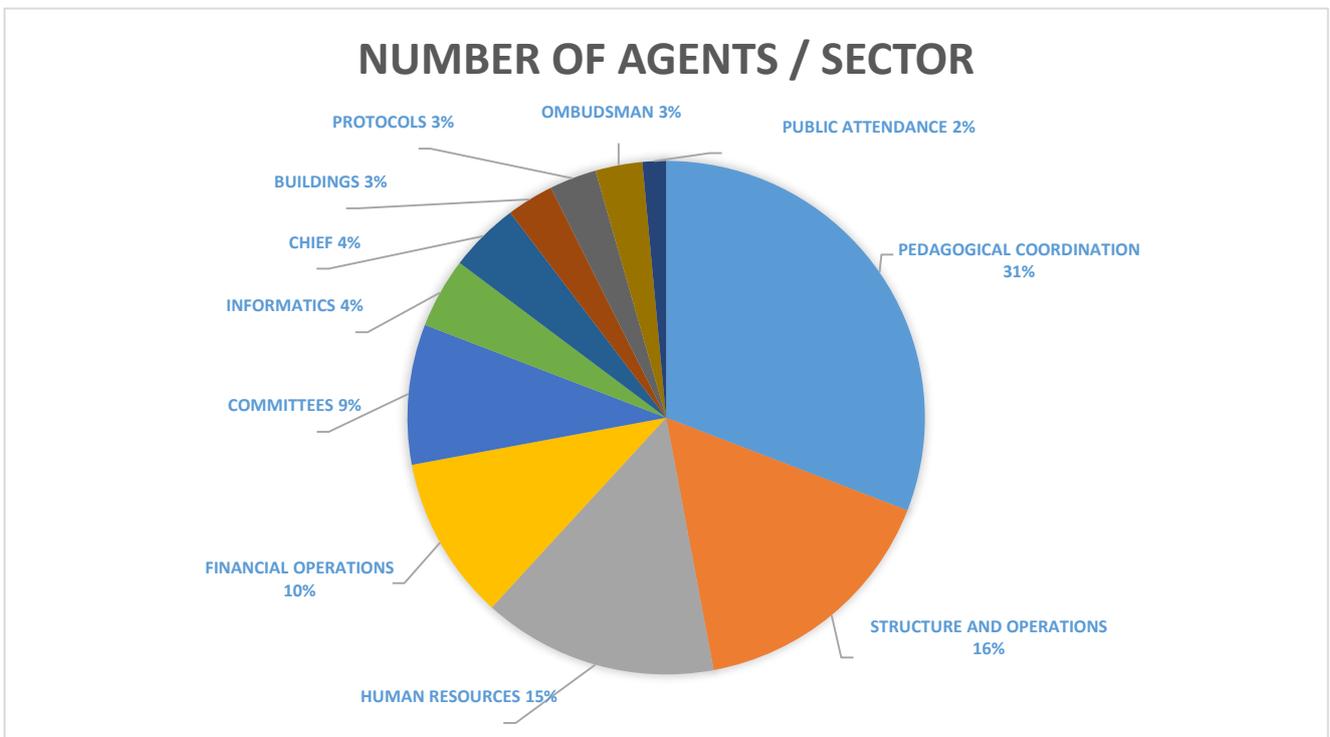

Figure 7. NRE of Francisco Beltrão, proportion of agents by work sector from figure 4.

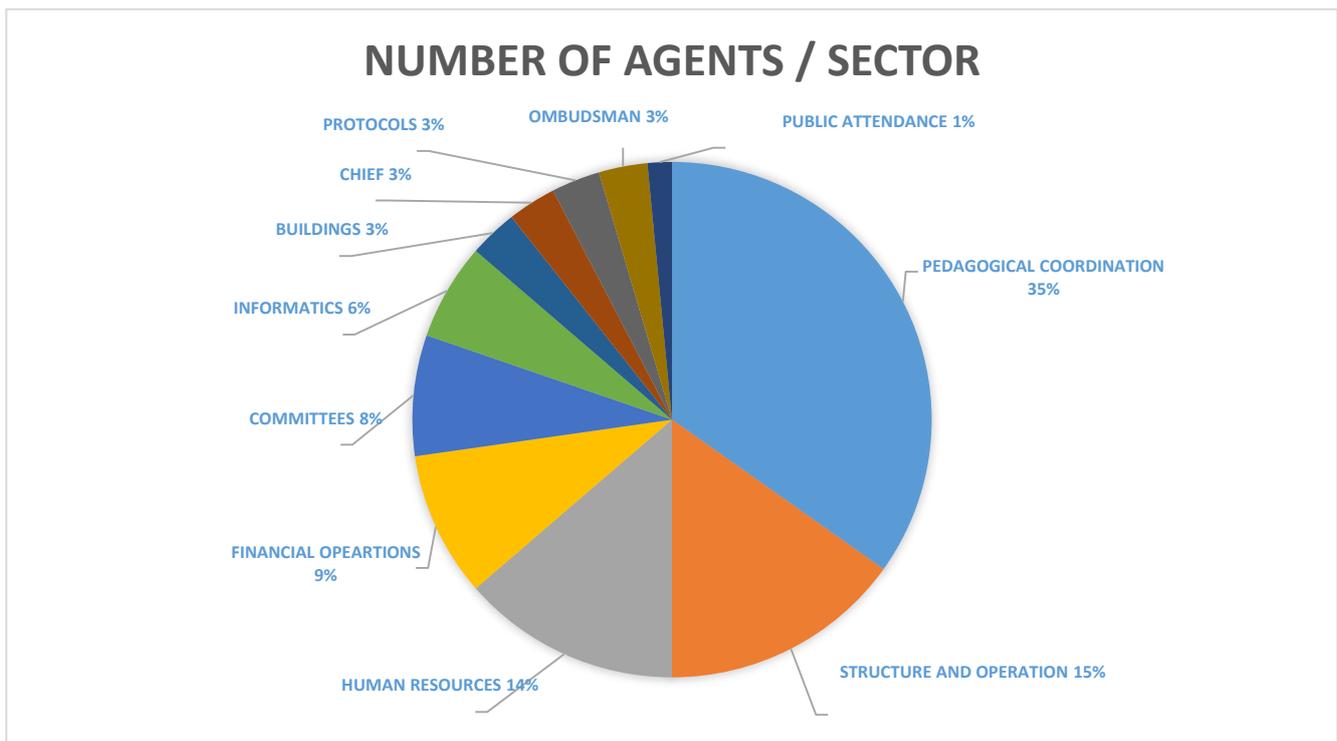

Figure 8. Pato Branco NRE, proportion between agents by work sector from Figure 5.

Several analyzes can be obtained from the data in figures 3-8 that seek to see which sector has the most work requests, which has the least agents, attempts to draw a correlation between number of agents and work requests. However, as mentioned in the methodology section, descriptive statistical analyzes such as those presented in figures 3-8 do not have sufficient explanatory power because the nature of administrative work does not always have discrete variables such as the beginning and end of work processes, which generates nonlinearity of events subjecting graphical analysis to management failure, but useful for monitoring as a final effect of managerial practices under investigation and experimentation.

The limit generated by nonlinear events in the descriptive statistical method is an important factor to consider as it represents that statistical methods can be useful only as an observation of the phenomenon in a steady state, and as mentioned in the methodology section, it should not be used for some types of event management or managerial strategy at risk that expected results will not be achieved.

### 3.1 An overview of administrative services frequency as a tool for focusing critical sectors

By correlating the work requests of each sample with the frequency of the administrative services classified as high, medium and low (see methodology section) in the qualitative research through interviews with the agents, the following work frequencies were obtained as shown in figures 9, 10 and 11. Note that the following figures differs from the previous ones due to the insertion of work frequencies in daily routine making possible a discretization of administrative routines

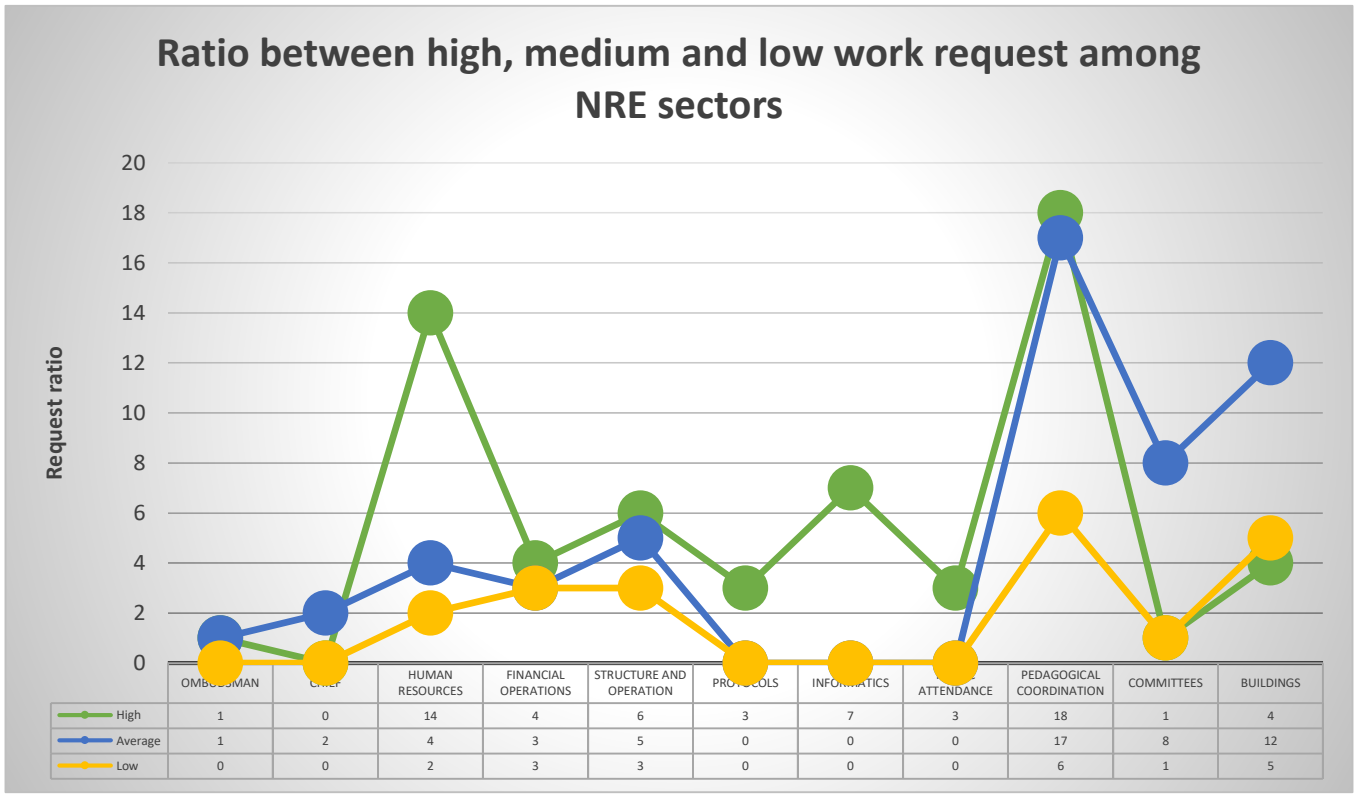

Figure 9. Dois Vizinhos NRE, frequency of work requests by sector from table 1.

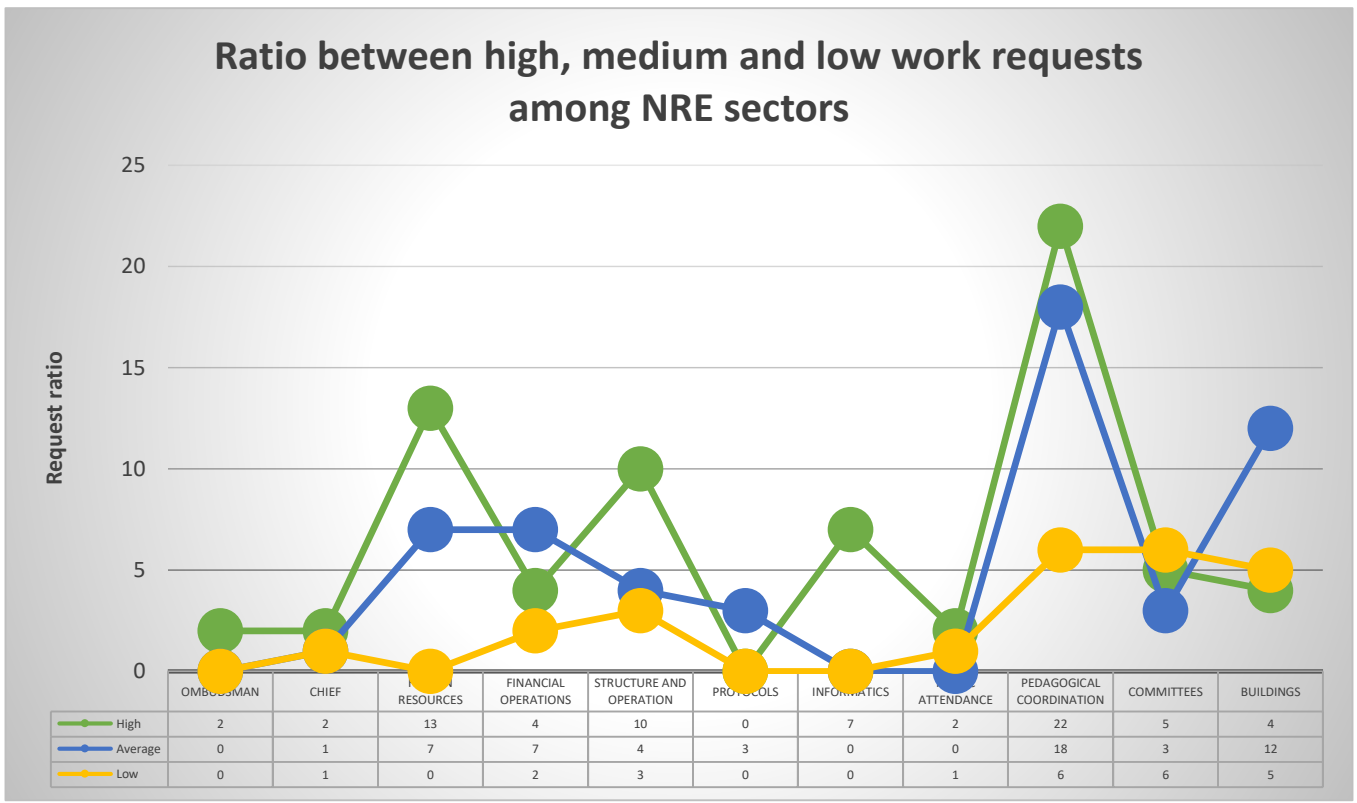

Figure 10. NRE Francisco Beltrão, frequency of work requests by sector from table 1.

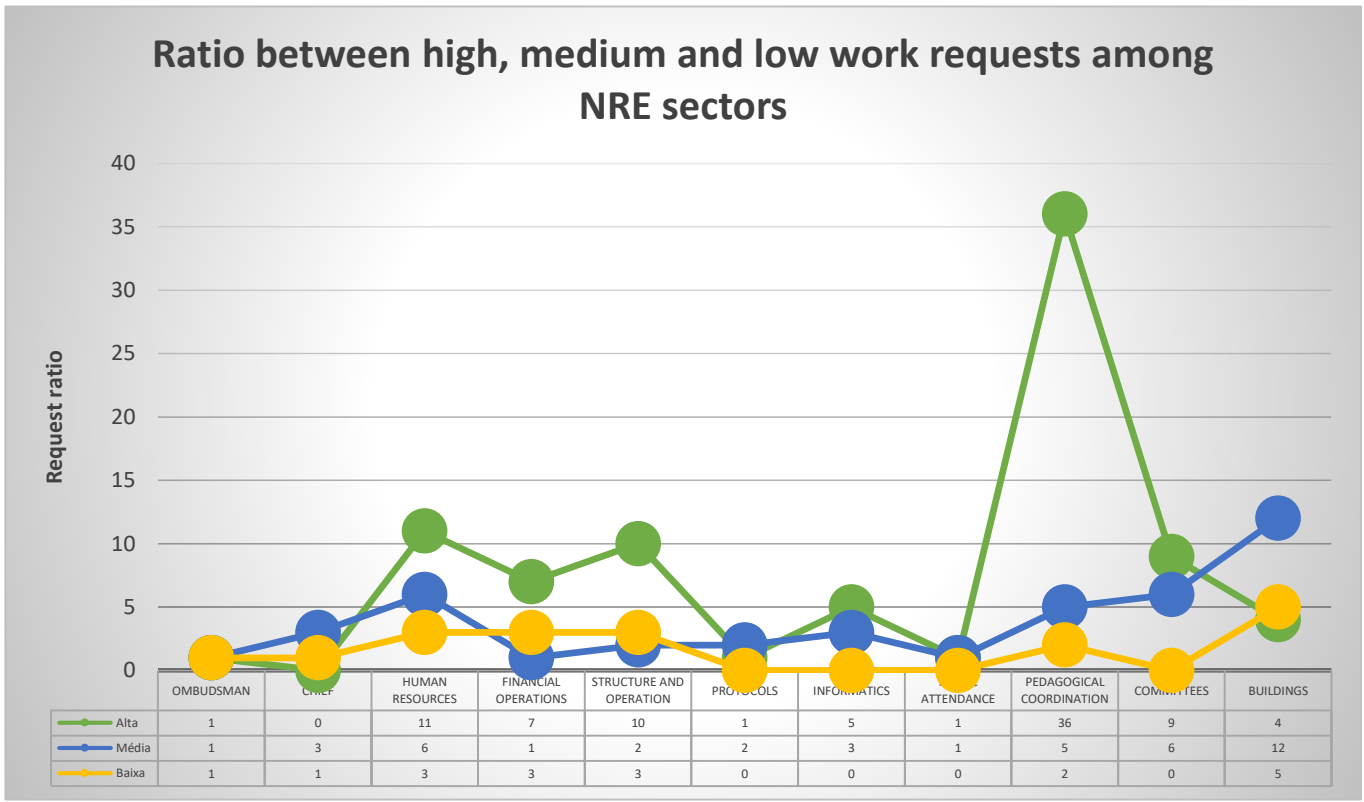

Figure 11. Pato Branco NRE, frequency of work requests by sector from table 1.

Figures 9, 10 and 11 do not present significant differences regarding work requests among the NRE. According to the statistical analysis, the Pato Branco NRE has the highest amount of high and low requests, while the Dois Vizinhos NRE has the highest average work requests.

A very significant aspect for all NREs analyzed is the fact that the building sector, in which an engineer performs the activities and methods resulting from the engineering activities, has a very low amount of human resources and a proportion of administrative services that in relation to other sectors, a possible high efficiency is identified regarding the proportion of human resources X administrative services. The work efficiency of the buildings sector is questioned, arguing that the work methods of the sector, although occurring in large quantities, are highly discretized and, as a result, many are already working digitally. When thinking about sectors whose administrative services did not reach sufficient discretization, not only is the possibility of computerizing the works limited, but their operationalization is conditioned to the agents in terms of information processing, in which there are several random effects that provide or not achieving the expected activity execution performance [13].

Importantly, even if the administrative services are not computerized, it does not reduce the efficiency with which it can be achieved, which, on the contrary, if it is not sufficiently discretized, the presence of continuous variables will certainly cause noise as the accuracy in performing a n activity.

An important aspect regarding the requests of work, is the potentiation of work when it occurs in an endo and exosectoral way. Looking at the sectoral administrative services of the NRE, we can see in figures 9, 10 and 11 that there is variation in the frequency and number of requests between the sectors.

By analyzing the data in figures 9, 10 and 11, it is possible to notice that some potential variable modifies the productivity of each NRE, which may be in the discretization of administrative services, computerization methods or in a variable that can be of great importance that is the of empowerment, in other words, the sharing of work between agents (see methodology section: endo and exosectoral concepts).

**3.2 Extraction of a nonlinear event indicator**

Figures 12, 13 and 14 show, as already mentioned in the methodology, a function that represents the input data sharing represented by the number of NRE agents and the endo and exosectoral collaborations and the output data as the existing administrative services represented by total by the number of public, municipal and private schools.

Note that the curvature of the function as well as the value of the coefficient of determination $R^2$, indicate, respectively, how much potential there is for action by agents and collaboration in proportion to the number of establishments. Note that Francisco Beltrão and Dois Vizinhos NREs are highly efficient at sharing work ($R^2 = 45\%$ and $R^2 = 44\%$) resulting in a ratio between input and output relatively higher than Pato Branco NRE ($R^2 = 33\%$).

This difference can be empirically evidenced by observing the actual numerical values of figures 12, 13 and 14.

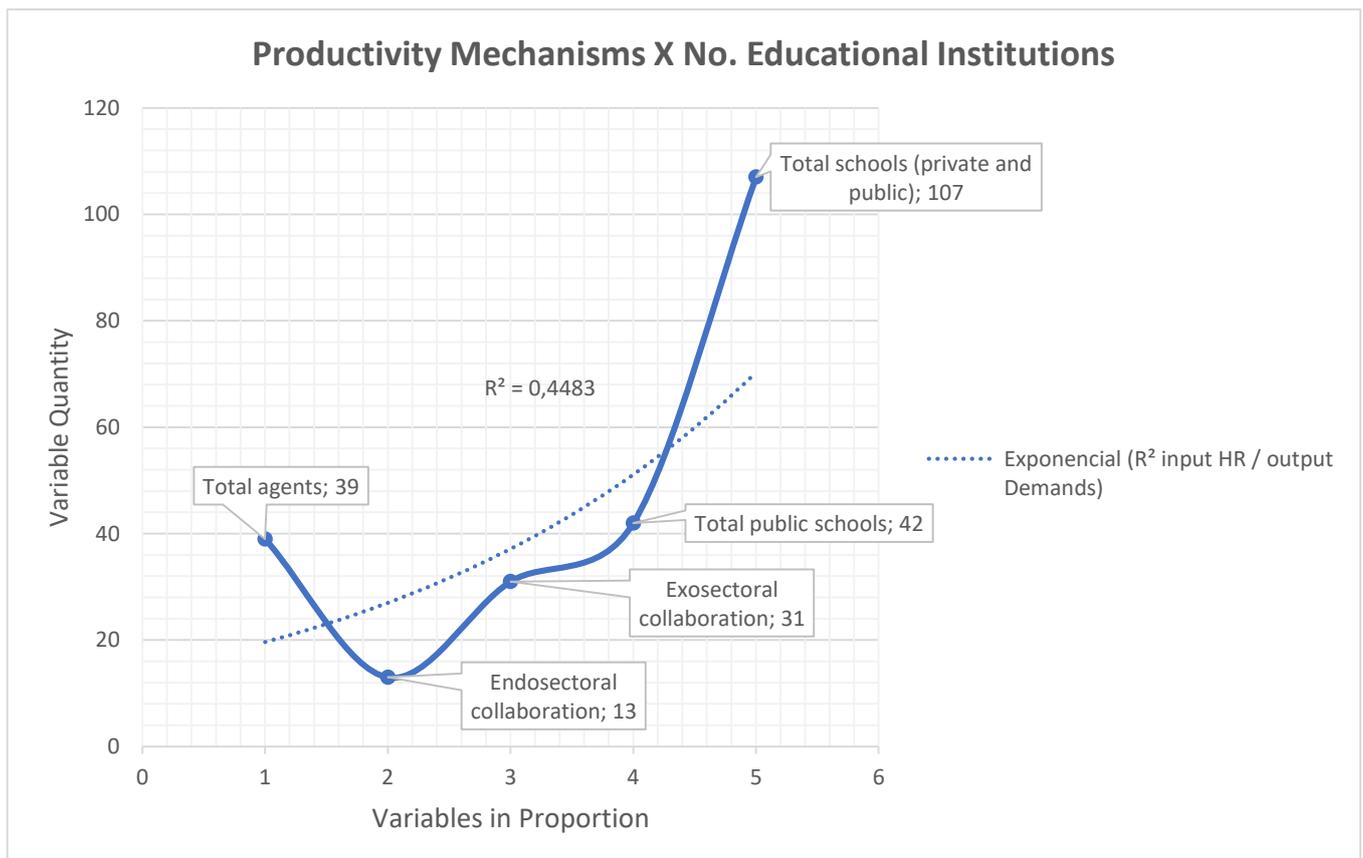

Figure 12. Dois Vizinhos NRE, endo and exosectoral labor relations, and proportion to labor demands.

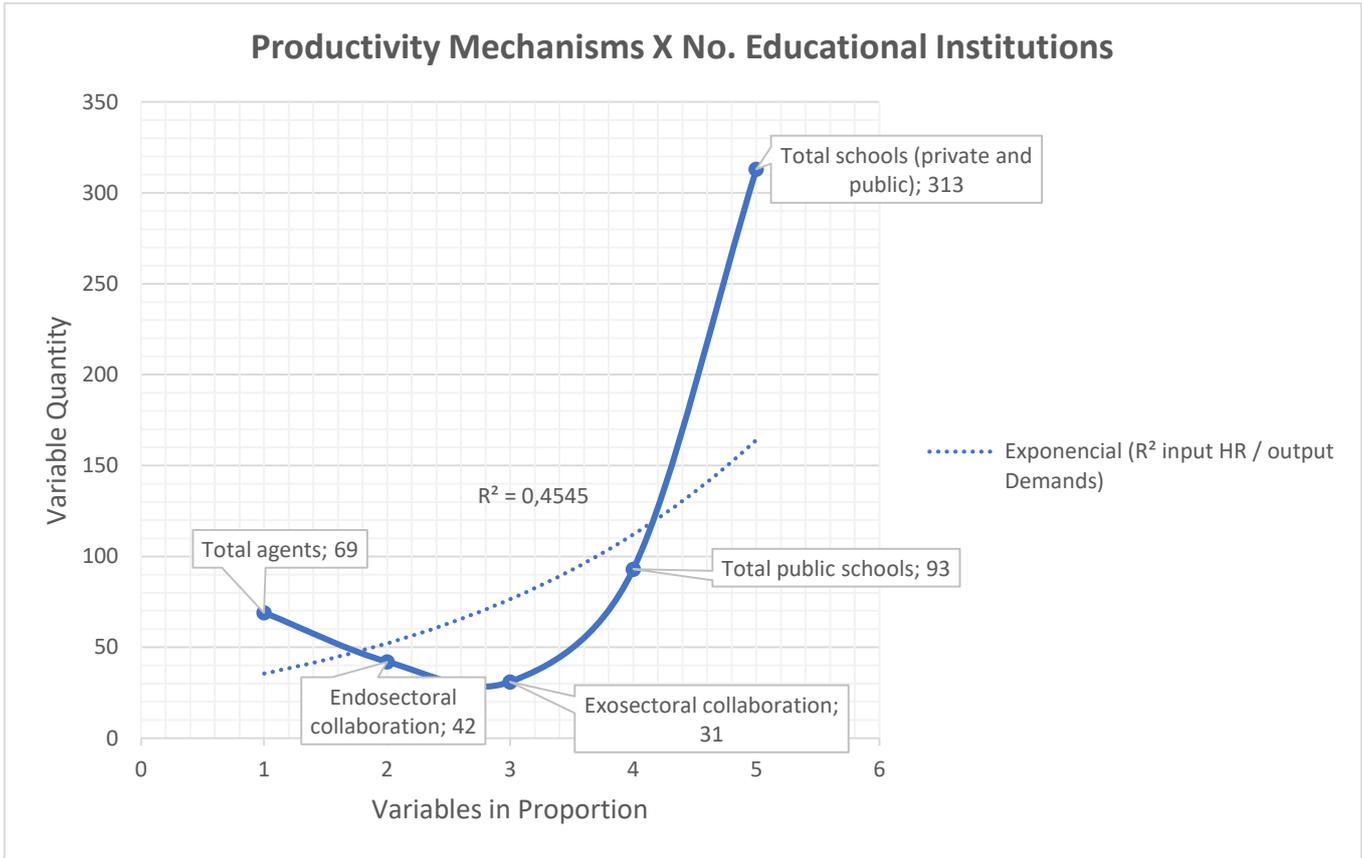

Figure 13. Francisco Beltrão NRE, endo and exosectoral labor relations and the proportion in relation to labor demands.

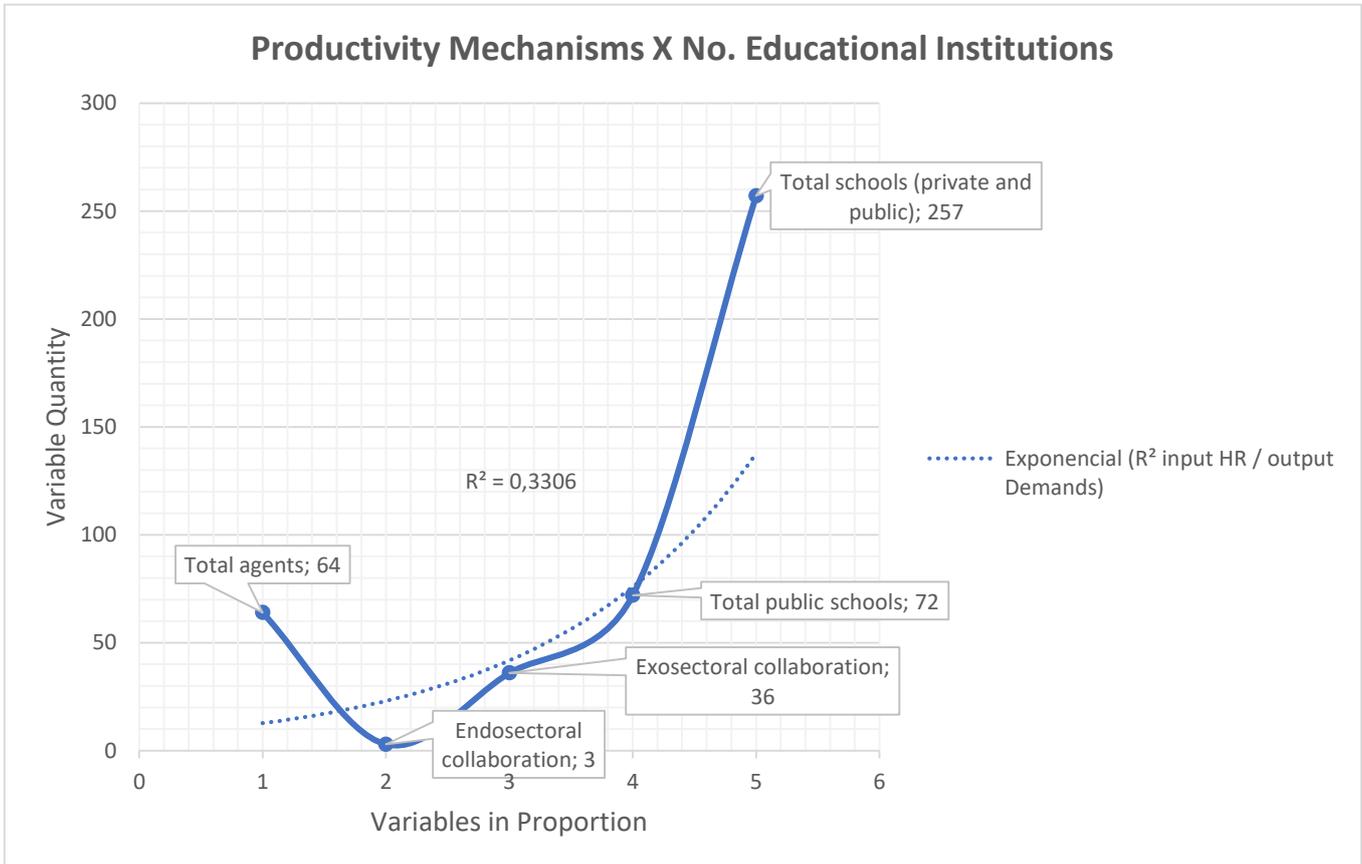

Figure 14. Pato Branco NRE, endo and exosectoral labor relations and the proportion in relation to labor demands.

## DISCUSSION

When thinking of a productivity indicator for the NRE, it was also defined as parameters all administrative actions that occur from the agent and / or educational policies that, regardless of how they are operationalized, assume as a final result the search for production efficiency from maximum limits originated by the number of agents and the number of educational establishments.

Observe that this method can be also applied to distinct sort of data distribution, that means in this research productivity was considered in the light of human resources and work requests represented by school establishments, but it can be designed for other structural purposes with other empirical variables. This flexibility makes possible to extend the use of the method to several other fields of work, not being exclusively conceived for administrative services.

In this sense, all administrative work, which includes all NRE flow of administrative services, assumes that the higher the level of information discretization, computerization of processes and improvements to achieve the desired performances, also higher production output, which as a rule, generates less need for human resources.

This is not to say that reducing human resources will improve productivity performance, but that operationalizations, requests and processing must have their performance improved before, so that at another time, the flow of system variables will converge to a lower number of human resources. This definition of administrative events allows system variables, whether linear or not, to assume maximum convergence at the fixed point defined by the number of educational establishments [15].

If, for example, a certain NRE has a high coefficient of determination, this means that in the proposed modeling as an indicator, labor demands through human resources are more proportional to the number of educational establishments than other NREs that do not have the same coefficient index. And in other words, it also reflects the conception of the phenomenon as efficient and effective with regard to discretization and / or computerization of workflows. But there is necessarily a balance in this statement, which concerns the number of educational establishments as well as agents as a whole of the system. Therefore, it is not enough to judge the coefficient of determination as unique data for fact analysis and later comparisons among sample quantities can be required to reach a more robust form of analysis searching for a ratio or other pattern that applies for variances within the collected samples.

However, although approximate, the coefficient value may indicate in the samples a concrete variation in how efficient human resources can be if managed in various ways.

Thus, by enabling the improvement of administrative processes, it is also possible the plural allocation of human resources to perform activities that have become faster and more practical to finalize their cycle of occurrence. In this sense, it is possible to see the event with the premise that the lower the workload of a given flow, the greater the power of human resources to manage multiple work executions, and on the contrary, the more primitive and precarious the workload, greater sectorization (individualization in the execution of activities), time spent to finish activities, reduction of the possibility of sharing work and response time to other dynamics dependent on events occurring in NRE.

It is therefore justified that an indicator that permeates the equation human resources, educational establishments and productivity also serves to observe how workflows are parsimoniously defined.

An example of a result with the use of the indicator is, for example, that as processes improve (discretization, computerization, ...), the number of work shares grows, the asymptotic effect of the exponential function curve decreases, and the coefficient of determination has its value increased. This indicates effectiveness in producing the administrative structure and allows new demands, if properly planned, to be passed on to the administrative unit to be executed with a high level of accuracy.

A final recommendation on the use of the indicator is that variables that influence NRE performances such as territorial extension, number of student enrollments, number of public agents of all types, among others, these quantifications do not incur the error regarding the proposal of the NRE indicator that measures productivity in terms of sharing among agents. Failure to use the method may occur if the data is falsified resulting in analysis errors or around the concept of sharing, even if an NRE indicates high productivity in this methodological sense, in case the agents present low accuracy in the execution of the activities, this will result in a loss of productivity and this event is outside the scope of the proposed indicator, but can be inductively identified by it.

**CONCLUSION**

The research presented a method to be used as an indicator of productivity in large-scale administrative jobs such as those found in the public federal, state or municipality administration.

The method assumes that for a nonlinear dynamics of the investigated administrative events, the best way to predict, manage and monitor the system and subsystems, it is necessary to conceive of the events as presenting exponential stability of Lyapunov as a mathematical modeling of the events.

Descriptive statistical methods that lead to a stationary observation of nonlinear systems were compared to the design of the same events as exponentially fixed-point dynamics, without defining a growth rate or reduction of the system power effect, but assuming this effect qualitatively as existing.

The forms of analysis were presented and in three samples, it was consistently obtained a differentiation between the productivity rate present in the systems (samples). The higher the input of human resources and the collaboration to perform activities in relation to the output of the number of educational establishments, the greater the proportionality of the exponential function used as well as the greater the curvature angle of the exponential function, indicating positive correspondence between the collected data and also converging empirically to the proposed methodology and the empirical evidence observed during the data collection phase.

**Acknowledgement:** Sincere thanks to Renan Veronesi Compagnoli for introducing Pareto's knowledge and methodology, employed empirically, which enabled the beginning of this research and the structure of thought.